\documentclass[12pt]{article}

\begin{document}

\hspace*{-15mm}\begin{minipage}{145mm}

\title{Schr\"odinger equation as the\\
      universal continuum limit of\\
      nonrelativistic coherent hopping\\ 
      on a cubic spatial lattice
\vspace{8mm}}
\author{Lutz Polley\\[5mm]Physics Department\\ 
                   Oldenburg University\\
		   D-26111 Oldenburg\vspace{20mm}}
\maketitle
\vspace*{12mm}
\thispagestyle{empty}
\abstract{The Schr\"odinger equation with scalar and vector 
potentials is the continuum limit of any coherent hopping process
(where position eigenstates superpose with neighbouring eigenstates
after a time step) whose hopping amplitudes are homogeneous in quadratic 
order of the inverse lattice spacing, inhomogeneous in first order,
and satisfying a summability condition with respect to 
higher-than-next neighbours.
\\[2mm]
}
\end{minipage}\newpage
\noindent
Discretization of space and time in the manner of
lattice gauge theory has greatly contributed to the intuitive  
understanding of fundamental processes \cite{Creutz}. For example,
differential equations like the Maxwell or continuity equations
appear as visualizable, geometric relations when they are written 
in terms of links, plaquettes, and (hyper)cubes. 
As I wish to show here, there is yet another fundamental differential law, 
the Schr\"odinger equation, which amounts to a rather intuitive statement
in terms of discretized space and quantum-mechanical superposition. 
The approach is similar to that of quantum lattice-gas models
\cite{Boghosian} used for efficient (quantum)computer simulation of the Schr\"odinger 
equation. Those models operate with unitary evolutions over a finite time step. 
It is then found that local algorithms require $2d$-component wave functions, 
where $d$ is the dimension of the spatial lattice. For single-component wave 
functions this implies that infinitesimal time steps are the only option,
if locality is to be maintained. 
In the present paper, therefore, time will be continuous as in 
Hamiltonian lattice gauge theory \cite{Kogut}.
 
To illustrate the point, consider the Schr\"odinger equation 
\begin{equation}   \label{1dimSchr}
i\frac{\partial\psi}{\partial t}=-\frac1{2m}\frac{\partial^2\psi}{\partial x^2} 
\end{equation}
on a 1-dimensional lattice with spacing $a$. A simple discretization of the second 
derivative is 
$$ 
 \frac{\partial^2\psi}{\partial x^2}  
                               = \frac{\psi(x+a,t)+\psi(x-a,t)-2\psi(x,t)}{a^2}
$$
After an infinitesimal step of time,
\begin{equation}   \label{1dimHop}
   \psi(x,t+{\rm d}t) = \psi(x,t) 
  + \frac{i{\rm d}t}{2ma^2}\{\psi(x+a,t)+\psi(x-a,t)-2\psi(x,t)\}
\end{equation}
Thus a wavefunction initially located on a lattice site will spread out to
next neighbours (quite in contrast to a continuum delta function which would
spread out uniformly to all of space) after a time-step ${\rm d}t$. 
This reflects the role of ``neighbours'' in the most basic concept of motion. 
Equation (\ref{1dimHop}) would therefore seem no less natural as an 
axiom than equation (\ref{1dimSchr}).  
Moreover, it is easy to argue that a quantum particle {\em must} move
by forming superpositions. This is because of the intriguingly simple 
axiom that an instantaneous particle position already determines the 
state vector. Hence, no information exists in a position state about the 
direction of hopping or motion.

The hopping amplitude $i{\rm d}t/2ma^2$ is certainly not an intuitive 
expression but can be taken, instead of the particle mass, as the 
{\it ab initio} parameter. 

In the following, hopping amplitudes are not restricted to any finite
number of neighbours, but are subjected to a summability condition, 
eq.\ (\ref{kappaKub}). It is further assumed that the hopping amplitudes 
realise 
the full translational and cubical symmetries of the lattice in 
${\cal O}(1/a^2)$ 
while any inhomogeneities in the hopping process are of ${\cal O}(1/a)$. 
At present I can only justify these assumptions by the fact that a 
renormalization
scheme exists in this case for the continuum limit $a\rightarrow0$.  
A further input is the conservation of probability. It will then be shown that 
any process of this kind results in the standard nonrelativistic Schr\"odinger 
equation in the continuum limit, with a vector potential (\ref{vecPot}) and a 
scalar potential (\ref{scalPot}).   

Consider a simple cubic lattice where $\vec{x}=a\vec{n}$ 
is the position vector of a site, $a$ is the lattice spacing, and 
$\vec{n}$ an integer vector. 
The most general equation for ``coherent hopping'', 
cast in the form of a differential equation with respect to time, is  
\begin{equation}    \label{SchrDiskrGen}
 \frac{{\rm d}}{{\rm d}t} \, \psi(\vec{x},t) 
         =  \sum_{\vec{n}} 
            \kappa(\vec{x},\vec{n}) \, \psi(\vec{x}+a\vec{n},t) 
\end{equation} 
The hopping parameters $\kappa(\vec{x},\vec{n})$ are complex numbers.
They could be time-dependent without changing the argument. 
Conservation of probability requires
\begin{equation}    \label{kappa-}
   \kappa(\vec{x}-a\vec{n},\vec{n}) = -\overline{\kappa(\vec{x},-\vec{n})} 
\end{equation}
An important case of reference is that of a free particle, characterized by
hopping parameters with the full symmetry of the lattice.
Then $\kappa(\vec{x},\vec{n}) = \kappa(\vec{n})$ because of translational 
invariance. Cubic symmetry implies 
$$
     \kappa(\vec{n}) = \kappa(-\vec{n})
$$ 
so that all $\kappa(\vec{n})$ are purely imaginary because of 
(\ref{kappa-}). Most importantly, the symmetry also implies 
$ \sum_{\vec{n}} \kappa(\vec{n}) \, n_i n_j \propto \delta_{ij}$. 
A convenient parametrization, incorporating the assumption about the order
of magnitude, is
\begin{equation}   \label{kappaKub}
  \sum_{\vec{n}} \kappa(\vec{n}) \, n_i n_j 
   =  \frac{i}{m a^2} ~ \delta_{ij} 
\end{equation}
The reduced parameter $m$ will be identified as the particle mass later on; 
the sign of $m$ is discussed in the Conclusions. 
In general, the sum in equation (\ref{kappaKub}) need not converge. 
Assuming convergence here is the basis for the nonrelativistic physics
as it emerges in the form of the Schr\"odinger equation in the continuum limit.

Deviations of the $\kappa(\vec{x},\vec{n})$ from the symmetry of the lattice
can be expressed through a complex-valued field $Z(\vec{x},\vec{n})$ defined,
in accordance with the ${\cal O}(1/a)$ assumption, by 
$$
  \kappa(\vec{x},\vec{n}) =  \kappa(\vec{n}) \, e^{i a Z(\vec{x},\vec{n})} 
$$
$Z(\vec{x},\vec{n})$ is more general than the U(1) lattice gauge potential
in two respects: it can have an imaginary part $\Im Z(\vec{x},\vec{n})$, 
and it is also defined for higher than next neighbours.
Nevertheless, standard procedures of lattice gauge theory suggest to consider
a ``lattice covariant derivative'' acting on any function $f(\vec{x})$ as
\begin{equation} \label {kovDif}
    D_{\vec{n}} f(\vec{x}) = 
   \frac1a \left( e^{i a Z(\vec{x},\vec{n})} f(\vec{x}+a\vec{n}) - f(\vec{x})
           \right) 
\end{equation}
For two covariant derivatives in succession one finds
\begin{eqnarray*}
 a^2 \, D_{-\vec{n}}D_{\vec{n}}\psi(\vec{x},t) & = &
  - e^{i a Z(\vec{x}, \vec{n})}\psi(\vec{x}+a\vec{n},t) 
  - e^{i a Z(\vec{x},-\vec{n})}\psi(\vec{x}-a\vec{n},t)  \\
& & + 
   \left( 1 + e^{i a Z(\vec{x},-\vec{n})} e^{i a Z(\vec{x}-a\vec{n},\vec{n})} 
         \right)  \psi(\vec{x},t) 
\end{eqnarray*}
The first two terms of this can be identified with hopping terms in 
(\ref{SchrDiskrGen}), and the last two with potential energy terms.  
As a consequence of (\ref{kappa-}),
$$
    D_{-\vec{n}} = \left( D_{\vec{n}} \right)^{\dag}
$$
Hence, equation (\ref{SchrDiskrGen}) can be rewritten as
\begin{equation}    \label{SchrKov}
   i \frac{{\rm d}}{{\rm d}t}~\psi(\vec{x},t) = 
   - \frac12 \sum_{\vec{n}} i a^2 \kappa(\vec{n})
   (D_{\vec{n}})^{\dag} D_{\vec{n}} \psi(\vec{x},t)
                              + W(\vec{x}) \psi(\vec{x},t)
\end{equation}
where, using (\ref{kappa-}) again,  
\begin{equation}    \label{Umagn} 
       W(\vec{x}) =  \frac12 \sum_{\vec{n}} i \kappa(\vec{n}) 
    \left( 1 + \exp \left( -2 a \Im Z(\vec{x},-\vec{n}) \right) \right)
\end{equation}
Now consider the continuum limit $a\rightarrow0$.
If the function $f$ in equation (\ref{kovDif}) is 
twice differentiable, it can be Taylor-expanded as 
$$ f(\vec{x}+a\vec{n}) = f(\vec{x}) 
  + a \vec{n}\cdot\vec{\nabla} f(\vec{x}) 
  + \frac{a^2}{2} \sum_{i,j=1}^3 n_i n_j ~
   \nabla_i \nabla_j f(\vec{x}+\vartheta a\vec{n})
$$
where $0<\vartheta(\vec{x},\vec{n})<1$. 
Expanding $\exp(iaZ(\vec{x},\vec{n}))$ also, one obtains
$$
   D_{\vec{n}} f(\vec{x}) =  \vec{n}\cdot\vec{\nabla}f(\vec{x}) 
              +  i Z(\vec{x},\vec{n}) \, f(\vec{x}) + {\cal O}(a) 
$$   $$
 (D_{\vec{n}})^{\dag} f(\vec{x}) = 
      - \vec{n}\cdot\vec{\nabla} f(\vec{x})
      + \overline{iZ(\vec{x},\vec{n})}f(\vec{x})  + {\cal O}(a)   
$$
The ${\cal O}(a)$ terms are omitted in the continuum limit, so that the
spatial derivative term of (\ref{SchrKov}) becomes
\begin{equation}   \label{DastD}
 - \frac12 \sum_{\vec{n}} i{a^2}\kappa(\vec{n})(D_{\vec{n}})^{\dag}D_{\vec{n}} 
 \, \psi(\vec{x},t)
 \quad = \hspace*{66mm}
\end{equation}
$$ 
\frac12 \sum_{\vec{n}} i{a^2}\kappa(\vec{n}) 
     \left( (\vec{n}\cdot\vec{\nabla})^2
  + i \Re Z \, \vec{n}\cdot\vec{\nabla}
  + i \vec{n}\cdot\vec{\nabla} \, \Re Z 
  - \vec{n}\cdot(\vec{\nabla}\Im  Z) - |Z|^2 \right) \, \psi(\vec{x},t)
$$
As it turns out, real and imaginary parts of $Z(\vec{x},\vec{n})$ affect 
the kinetic energy very differently. 
The $\Re Z$ terms can be interpreted as vector potential terms, if one defines
\begin{equation}    \label{vecPot} 
   \vec{A}(\vec{x}) =  i m a^2  
   \sum_{\vec{n}} \kappa(\vec{n}) \, \vec{n} \,
                     \Re  Z(\vec{x},\vec{n}) 
\end{equation}
The $\Im Z$ terms only contribute to the scalar potential.
Another contribution to the latter is produced by
completing the square with respect to $\vec{A}(\vec{x})$.
Thus, starting from (\ref{SchrKov})
and using (\ref{DastD}), (\ref{kappaKub}), and (\ref{Umagn}), 
one arrives at the general, nonrelativistic Schr\"odinger equation 
\begin{equation}   \label{SchrMag}
    i\frac{\partial}{\partial t}\psi(\vec{x},t) = 
    \frac{1}{2m}  
  \left(i\vec{\nabla} + \vec{A}(\vec{x})\right)^2  
  \psi(\vec{x},t) + U(\vec{x}) \psi(\vec{x},t)  
\end{equation}
where 
\begin{equation}    \label{scalPot} 
   U(\vec{x}) = W(\vec{x}) 
 - \frac12 \sum_{\vec{n}} i{a^2}\kappa(\vec{n}) \left(
   \vec{n}\cdot\vec{\nabla}\Im Z(\vec{x},\vec{n})
  + | Z(\vec{x},\vec{n}) |^2 \right) 
 - \frac{\vec{A}(\vec{x})^2}{2m}
\end{equation}
The scalar potential $U(\vec{x})$ can be given any value desired 
by adjusting $\kappa(\vec{n})$ and $Z(\vec{x},\vec{n})$ with $\vec{n}=0$.
These on-site hopping parameters neither affect the mass 
in eq.\ (\ref{kappaKub}) nor the vector potential in eq.\ (\ref{vecPot}).
But they do contribute a divergent term to $W(\vec{x})$ in (\ref{Umagn}), 
appropriate for additive renormalization, 
and a further, non-leading term to $U(\vec{x})$ in (\ref{scalPot}). 

In conclusion, one can derive the general form of the Schr\"odinger 
equation using:
(i) the absence of motional information from a position eigenstate, 
(ii) the superposition principle for undecided alternatives, 
(iii) the statistical interpretation of wave functions,
(iv) an assumption about the summability of hopping amplitudes
(nonrelativistic condition), and
(v) an assumption about the dependency of the hopping parameters
on the lattice spacing (renormalization scheme).

The sign of the parameter $m$ in equation (\ref{kappaKub})
was assumed to be positive. This is a matter of convention only. The 
general implication of ``coherent hopping'' is that the kinetic 
energy of a free particle is either always positive or always negative. 

For the definition of the mass parameter in (\ref{kappaKub})
it was essential that a free particle
find identical hopping conditions on every site of the lattice. 
But this also characterizes the lattice as a cartesian 
coordinate system. A generalized hopping scenario may thus explain why 
cartesian coordinates play such a preferred role 
in a wide range of quantum systems \cite{CristLee}.   

A point worth emphasizing is that the link variables of U(1) lattice gauge 
theory can be extended from phase factors to arbitrary complex numbers
without changing the form of the Schr\"odinger equation---the 
extensions only contribute to the potential energy. 
Preliminary studies of the quantized U(1) gauge field
indicate \cite{Polley} that analogous terms,
referring to hopping in the U(1) configuration space,
are responsible for the magnetic field energy.

\subsubsection*{Acknowledgments}

I wish to thank  B. M. Boghosian, E. Mendel, M. Nest, and J. Pade for important 
suggestions.


\begin{thebibliography}{1}

\bibitem{Creutz} K. Wilson, Phys.\ Rev.\ D 10 (1974) 2445; 
                 M. Creutz, Quarks, gluons and lattices, 
                 Cambridge 1983. 

\bibitem{Boghosian} B. Boghosian, W. Taylor, Phys.\ Rev.\ E 57 (1998) 54; 
                 Physica D 120 (1998) 30; Int.\ J. Mod.\ Phys.\ 8 (1997) 705. 

\bibitem{Kogut}  J. Kogut, L. Susskind, Phys.\ Rev.\ D 11 (1975) 395.  

\bibitem{CristLee}  N. H. Christ, T. D. Lee, Phys.\ Rev.\ D 22 (1980) 939.

\bibitem{Polley} L. Polley, work in progress.

\end{thebibliography}
\end{document}